\documentclass[journal,11pt,draftclsnofoot,onecolumn]{IEEEtran}
\usepackage{cite}
\usepackage{epsfig}
\usepackage{graphicx}
\usepackage{url}
\usepackage{amsfonts}
\usepackage{amsmath,bm}
\usepackage{amssymb}
\usepackage{multirow}
\usepackage{multicol}
\usepackage{times}
\usepackage{psfrag}
\usepackage{subfigure}
\usepackage{stfloats}
\usepackage{footnote}
\usepackage{array}
\usepackage{booktabs, threeparttable}
\usepackage{setspace}

\allowdisplaybreaks

\usepackage{algorithm}
\usepackage{algorithmic}
\usepackage[T1]{fontenc}

\begin{document}

\title{Achieve Sustainable Ultra-Dense Heterogeneous Networks for 5G}

\author{Jianping An,~\IEEEmembership{Member,~IEEE}, Kai Yang,~\IEEEmembership{Member,~IEEE}, Jinsong Wu,~\IEEEmembership{Senior Member,~IEEE}, Neng Ye, Song Guo,~\IEEEmembership{Senior Member,~IEEE}, and Zhifang Liao

\thanks{This work was supported by the National Natural Science Foundation of China under Grant 61501028 and Grant 61771054. \textit{(Corresponding author: Kai Yang.)}}

\thanks{J. An, K. Yang, and N. Ye are with the School of Information and Electronics, Beijing Institute of Technology, Beijing, China, and also with Beijing Key Laboratory of Fractional Signals and Systems, Beijing, China (email: yangkai@ieee.org).}
\thanks{J. Wu is with the Department of Electrical Engineering, Universidad de Chile, Santiago, Chile.}
\thanks{S. Guo is with the Department of Computing, Hong Kong Polytechnic University, Hong Kong, China.}
\thanks{Z. Liao is with the School of Software, Central South University, Changsha, China.}}

\maketitle

\begin{abstract}
Due to the exponentially increased demands of mobile data traffic, e.g., a 1000-fold increase in traffic demand from 4G to 5G, network densification is considered as a key mechanism in the evolution of cellular networks, and ultra-dense heterogeneous network (UDHN) is a promising technique to meet the requirements of explosive data traffic in 5G networks. In the UDHN, base station is brought closer and closer to users through densely deploying small cells, which would result in extremely high spectral efficiency and energy efficiency. In this article, we first present a potential network architecture for the UDHN, and then propose a generalized orthogonal/non-orthogonal random access scheme to improve the network efficiency while reducing the signaling overhead. Simulation results demonstrate the effectiveness of the proposed scheme. 
Finally, we present some of the key challenges of the UDHN.
\end{abstract}

\IEEEpeerreviewmaketitle

\section{Introduction}
Fueled by the proliferation of mobile devices and applications, mobile data traffic increases exponentially in the past decade, and this trend will certainly continue and accelerate even further in the future \cite{IEEECM2016_MCUDN}. It has been reported that the mobile device density (the device number per unit area) and traffic density (the traffic volume per unit area) will rise to several million per $km^2$ and tens of Tera bits per second per $km^2$ in local hotspot areas, respectively \cite{IMT2020}. These explosions of mobile device and traffic would impose new challenges to future cellular networks. 

To tackle the 1000x mobile traffic challenge and to accommodate the massive devices, ultra-dense heterogeneous networks (UDHN) have been identified as a key mechanism to handle orders of magnitude increase in the data traffic volume \cite{IEEECM2016_MCUDN,IMT2020,IEEECM2014_NetDensif,IEEECM2015_SE_EE_UDN}. The UDHN refers to the idea of densifying the cellular networks with a very high network densification, including both the mobile device densification and base station (BS) densification, where the density of BS may exceed that of mobile device. As a result, the UDHN would potentially allow for orders of magnitude improvement in both spectral efficiency (SE) and energy efficiency (EE) \cite{IEEECM2015_SE_EE_UDN}.

The more traffic generated, the more BSs will be needed to serve the devices. The evolution of the cellular networks, i.e., from 1G to 5G, can be viewed as the process of network densification to some extent. In 1G networks, the cell radius is around 10 miles, and cell splitting may occur, where the radio coverage of one cell is partitioned into two or more new smaller cells to mitigate path-loss and to support more devices. In 2G networks, the typical radius of macrocell varies from a couple of hundred meters to several kilometres. Besides cell splitting, small cells, which are low-powered radio access nodes that have a range of tens of meters to 1 or 2 kilometers, are introduced in 2G networks to offload traffic from macrocells. In 3G and 4G networks, small cells are further popular and viewed as a vital element for boosting network throughput and balancing traffic load. Deploying more and more small cells to serve a certain geographical area leads to smaller and smaller cell coverage areas, and thereby achieving cell splitting gains. With small cells being deployed, homogeneous cellular networks become heterogeneous because of difference in transmit power and coverage area of BSs. In 5G networks, the density of small cells will be further increased compared with that in 4G networks to guarantee seamless coverage and high data rate. As a result, the UDHN is emerging as one of the core characteristics of 5G cellular networks \cite{IEEEWC2016_5GUDCN}. However, the performance of the UDHN doesnot increase monotonously with BS density, and it is limited by the inter-cell interference and the fronthaul and backhaul network capacity \cite{IEEEWC2016_Fronthauling,IEEEWC2016_5GUDCN}. The full benefits of network densification can be realized only if it is complemented by advanced transceivers capable of interference cancellation and backhaul densification \cite{IEEECM2014_NetDensif}.

In this article, we first present one potential network architecture for 5G UDHN, and then investigate the random access problem in 5G UDHN. Here, we focus on the machine type communication (MTC) devices, whose density is much higher than that of human type communication (HTC) devices, and propose a random access scheme for MTC devices to improve the network efficiency while reducing the signaling overhead. Finally, we analyze the challenges of 5G UDHN. 

\section{Network Architecture}

\begin{figure}
    \centering
    \includegraphics[scale=1]{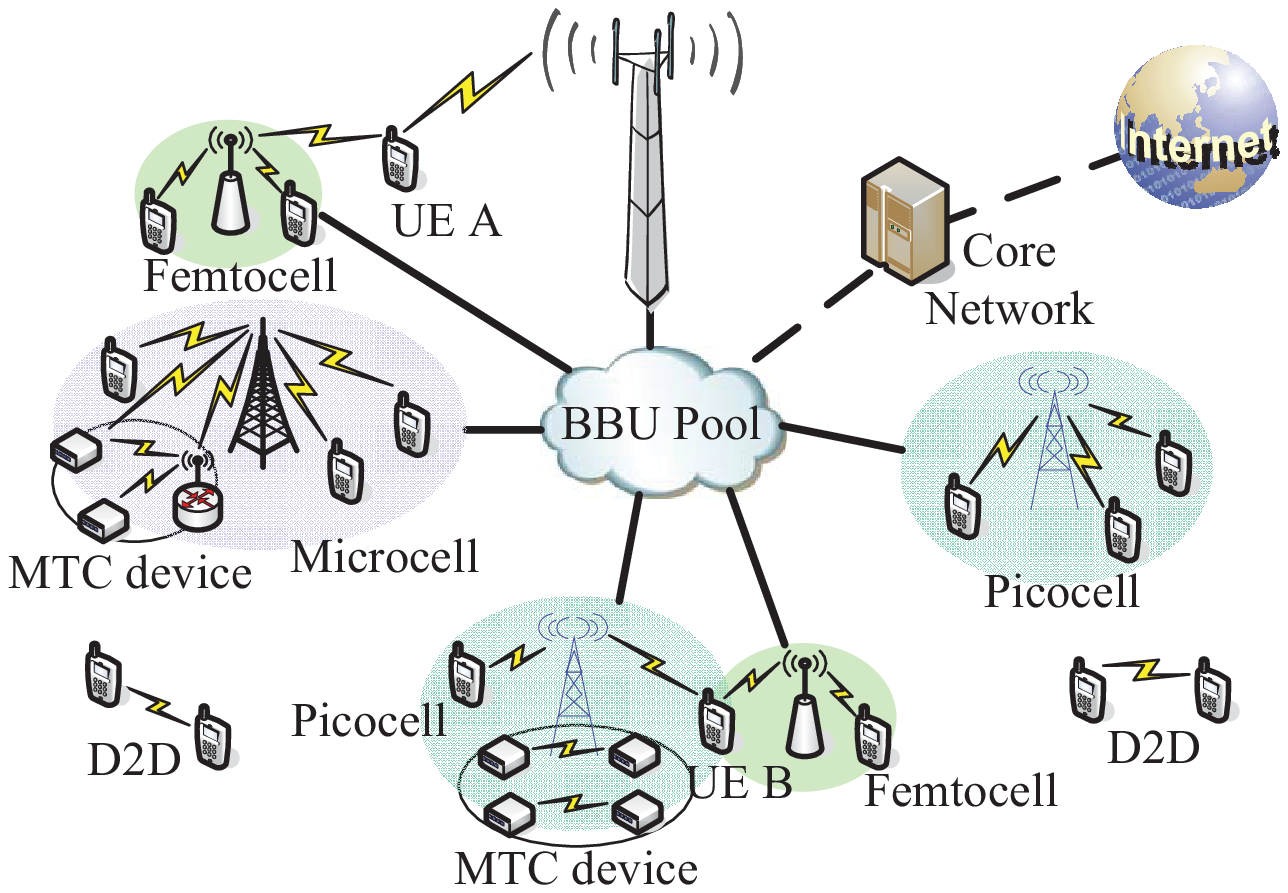}
    \caption{The proposed potential network architecture for 5G ultra-dense heterogeneous networks.}
    \label{Fig_Architecture}
\end{figure}

The current LTE heterogeneous networks are mainly designed for mobile broadband scenario. In addition, the communications between different cells are limited by the existing Un and Uu interface, which may not satisfy the demand of growing traffic. To meet the 5G requirements on explosive growth in traffic volumes and unprecedented increase in connected mobile devices, dramatic changes in the design of network architecture are required. As a result, we propose a potential network architecture for the UDHN, as shown in Fig. \ref{Fig_Architecture}, where macrocells, microcells, picocells, and femtocells are  connected via the cloud radio access network (C-RAN) structure to facilitate the collaboration among cells. The proposed UDHN can be viewed as an evolution of the current LTE heterogeneous networks, and is compatible with the current LTE architecture.

\subsection{Multi-Tier Heterogeneous Networks}
According to the trend in cellular networks evolution, 5G networks will be a heterogeneous one consisting of macrocells along with a large number of small cells, device-to-device (D2D) pairs, and MTC devices based communication tiers \cite{IEEECST2016N5G}. Each tier has a different size with a different transmit power level. Typically, macrocell is covered by a high power node with a large radius to provide wide area coverage for the remote and rural areas, whereas the main purpose of small cells, which encompass microcells, picocells, and femtocells, is to enhance network capacity in hot spots with very dense services through offloading traffic load from the macrocells and increasing the frequency reuse factor. Compared with microcell and picocell, which are deployed by operator, femtocell is typically in the mode of plug and play, and any consumer can install a femtocell at home. Due to the private property, consumer needs to declare the femtocell access control protocol, namely, open access control, where any nearby users can access the femtocell, or closed access control, where only the pre-registered users are allowed to access the femtocell.

Due to the ultra-densities of BSs and users, a tractable model for the UDHN is desired to facilitate the network performance analysis compared to complex system-level simulations. Fortunately, we can model multi-tier UDHN based on stochastic geometry, where we model the locations of BSs of the same tier as a homogeneous Poisson point process (HPPP) of density $\lambda$. Different tier is with different $\lambda$ to characterize its density property. It is easy to theoretically analyze the performance of multi-tier UDHN based on HPPP model by using Palm measure. However, the locations of the macro BSs are usually dominated by the network planning, and are not random in general. To model the location of macro BS more appropriate, we can resort to Matern hard core Poisson point process, whose repulsive nature makes it a suitable candidate to model macro BSs in cellular networks. Furthermore, we can use Poisson cluster process to characterize the clustering property of small cells, as small cells are usually clustered around highly populated areas.

\subsubsection{Device-to-Device Communications}
To take advantage of the physical proximity of communicating devices, such as increasing resource utilization, improving cellular coverage, and offloading traffic load from BSs, D2D communications are proposed as a key component for future cellular networks, as shown in Fig. \ref{Fig_Architecture}. Two D2D UEs can communicate with each other using the licensed radio resources in three modes, i.e., cellular mode, overlay mode, and underlay mode, to coordinate the cellular and D2D links and to manage the interference while remaining control under the network. Compared with the underlay mode, the overlay mode is with the advantage in facilitating the challenging interference management. In underlay mode, resource sharing between D2D pair and cellular UE incurs interference, which is usually viewed as an obstacle to cellular networks. 

To realize the direct communications between two D2D UEs, the transceiver architecture of current UE shall be changed, since LTE adopts orthogonal frequency division multiplexing (OFDM) and single-carrier frequency division multiple access (SC-FDMA) in the downlink and uplink, respectively. As a result, we can choose either OFDM or SC-FDMA for D2D communications. One simple solution is to add IDFT module in the UE receive chain for receiving SC-FMDA signal or bypass the DFT module in the UE transmit chain for transmitting OFDM signal. The D2D communication ranges with different transceiver architectures are shown in Table \ref{Talbe_D2Drange}, where we observe that the range improvement of SC-FDMA-based architecture is about 10$\sim$15\% and 20$\sim$50\% for indoor and outdoor scenarios, respectively, compared with OFDM-based architecture. This is due to the fact that the peak-to-average power ratio of OFDM is higher than that of SC-FDMA, which means that the power back-off of SC-FDMA is lower than that of OFDM.

\begin{table}[!t]
	\centering\caption{D2D Communication Ranges (meters): OFDM vs. SC-FDMA.}\label{Talbe_D2Drange}
	\begin{threeparttable} 
		\begin{tabular}{cccccccccc}
			\toprule 
			\multicolumn{2}{c}{Modulation Scheme}  &\multicolumn{2}{c}{QPSK} & \multicolumn{2}{c}{16QAM} &	\multicolumn{4}{c}{64QAM} 	\\
			\multicolumn{2}{c}{Code Rate} & 1/2 & 3/4 & 1/2 & 3/4& 1/2	& 2/3	& 3/4 & 5/6	\\\hline\noalign{\smallskip}
			\multirow{2}{*}{Indoor Scenario}&SC-FDMA	&64.7	&58.8	&54.6	&48.9	&48.7	&44.2	&42		&39.8\\
			&OFDM	&56.4	&50.6	&49.5	&43.9	&45		&40.5	&38.4	&36.2\\\hline\noalign{\smallskip}
			\multirow{2}{*}{Outdoor Scenario}&SC-FDMA	&378	&283.5	&231.1&	173.3&	171.8	&136.5&	121.6&	108.4\\
			&OFDM	&252.6	&189.5	&178.9	&134.1	&142.1	&112.8&	100.6	&89.6\\\bottomrule
		\end{tabular}
	\end{threeparttable}
\end{table}

\subsubsection{Machine Type Communications}
Compared with HTCs, one of the key characteristics of MTCs is that it involves a potentially huge number of power-limited devices, where each connected device transmits low volume of non-delay sensitive data with a certain period. Since the current cellular networks are designed for human type applications, the operator should accommodate their networks to support MTCs. It is noticed that handling a huge amount of MTC devices causes problems in connection establishment and radio resource allocation \cite{IEEECM2015MTC}, and significant congestion problem will be risen due to simultaneous signaling or data transmissions from massive MTC devices. 

Before establishing a radio resource control (RRC) connection for an MTC device, a random access procedure shall be initialized for the MTC device. With a large number of MTC devices performing the random access procedure simultaneously, the succession probability of access for each device would degrade seriously. In order to mitigate the congestion in random access procedure, a simple solution is to assign new dedicated resource for the access of MTC devices based on the traffic conditions and the number of MTC devices. In this article, we will propose a random access scheme for massive MTC devices to mitigate the congestion and to improve the network performance later.

\subsection{Downlink and Uplink Decoupling}
Compared with the traditional coupled downlink-uplink association strategy in current cellular networks, decoupling of downlink and uplink is inspired to balance the traffic load between downlink and uplink in multi-tier UDHN, where each user may be associated with different BSs in the downlink and uplink directions and the downlink and uplink communication sessions are treated as two separated entities. From the perspective of optimization, the region of coupled association is a subset of that of decoupled association, which indicates that a well designed decoupled association strategy principally outperforms coupled one in terms of network throughput, SE, and EE \cite{IEEECM2016Decouple}.

Downlink and uplink decoupling (DUDe) can be viewed as an evolution of the biased coupled association, where the small cell range is extended through the use of a positive cell selection offset to the largest received downlink power. Compared with biased coupled association, the gains of DUDe are substantial. A simple access rule of DUDe is that users connect to the nearest BS in uplink and access to the BS with the strongest received downlink signal in downlink. This is much more beneficial for MTC applications, whose uplink traffic load is usually much higher than downlink traffic load. 

Compared with coupled association, DUDe results in new challenges from the perspective of network design, where the logical and physical channels are much easier to design and operate for coupled association \cite{IEEECM2016Decouple}. Furthermore, decoupling the downlink and uplink requires a strict synchronization and signalling connectivity between the two association BSs, which can be easily implemented with C-RAN. In C-RAN, two BSs can exchange information directly to perform call admission, handover, and resource assignment, which could facilitate the DUDe. As the C-RAN implements the protocol stack in a centralized fashion, it is preferred to employ a centralized fashion for DUDe in our proposed UDHN with C-RAN architecture.

\subsection{Cloud Radio Access Network}
Since ultra-dense networks meet new challenges of densely deploying small cells due to the significant capital expenditure (CAPEX) and operating expenditure (OPEX), C-RAN is viewed as a promising paradigm for 5G UDHN, as shown in Fig. \ref{Fig_Architecture}. C-RAN integrates cloud computing into radio access networks to virtualize the functionalities of BSs, and it may be viewed as an evolution of the distributed BS. In distributed BS, the radio function unit, i.e., remote radio head (RRH), is separated from the digital function unit, i.e., baseband unit (BBU). In C-RAN, BBUs are brought together to construct a BBU pool, and RRHs are distributed deployed to provide wide coverage and high data rate \cite{IEEEWu,IEEECM2016CRAN}. A number of RRHs are connected a BBU pool via high bandwidth and low latency links, e.g., optical fiber or microwave connections, using the open base station architecture initiative (OBSAI) or common public radio interface (CPRI) standard \cite{IEEECST2016CRAN}. As the BBU pool is responsible for all the baseband signal processing, C-RAN could facilitate the collaboration among cells, such as coordinated multipoint (CoMP) and interference mitigation.

In Fig. \ref{Fig_Architecture}, all the macrocell, microcell, picocell, and femtocell can be served by RRHs, which transmit radio frequency (RF) signals to users in the downlink and forward the baseband signals to BBU pool in the uplink. As most of the functions of the protocol stack are conducted in BBU pool, RRHs could be relatively simple and cheap with PHY layer functions only. Compared to stand-alone BS, which conducts the entire protocol stack and has its own backhaul, battery, air condition, and so on, RRH is a pure RF unit, and can be easily deployed to extend the coverage and to improve the capacity in a cost-efficient manner. Compared to traditional network architecture with stand-alone BSs, the baseband processing resources are shared among RRHs through cloud computing in BBU pool, and fewer BBUs are needed in C-RAN to cover the same geographic area. It is reported that one BBU pool could serve up to one thousand RRHs to cover a geographic area with a radius of tens of kilometers. Due to the cloud computing property of C-RAN, it could also effectively solve the challenge of traffic load fluctuation of each cell caused by the user mobility, and saves the processing resources and power at idle times. 

Because of the densely deploying of inexpensive RRHs and processing resources sharing among cells, C-RAN architecture could significantly reduce the CAPEX and OPEX compared to current network architecture with stand-alone BSs and distributed BSs. Furthermore, due to the scalability of C-RAN, it is easy to add new BBUs in the BBU pool to improve the processing ability and to deploy new RRHs to enhance the network coverage, thereby facilitating system roll out and network maintenance.

However, C-RAN architecture brings huge overhead on the fronthaul links between RRHs and BBU pool, and the capacity of fronthaul has a crucial influence on the performance of C-RAN. The traffic load on the fronthaul is proportional to the aggregate volume of all the users, and it is 50 times higher than that on the backhaul. For example, the CPRI data rate per carrier of current TD-LTE is 10 Gbps. Hence, the fronthaul capacity is a bottleneck of C-RAN. To tackle this problem, fronthaul compression is necessary, where RRHs and BBU pool compress the baseband signals in uplink and downlink respectively. With the continued user densification in 5G UDHN, the impact of the constrained fronthaul will become more and more serious, and joint design from both the wireless and transport perspectives is required for the advanced fronthaul. With constrained fronthaul, fronthaul-aware resource allocation schemes are also required to alleviate the fronthaul constraints.

\section{Random Access}\label{RA}
\begin{figure}
    \centering
    \includegraphics[scale=1]{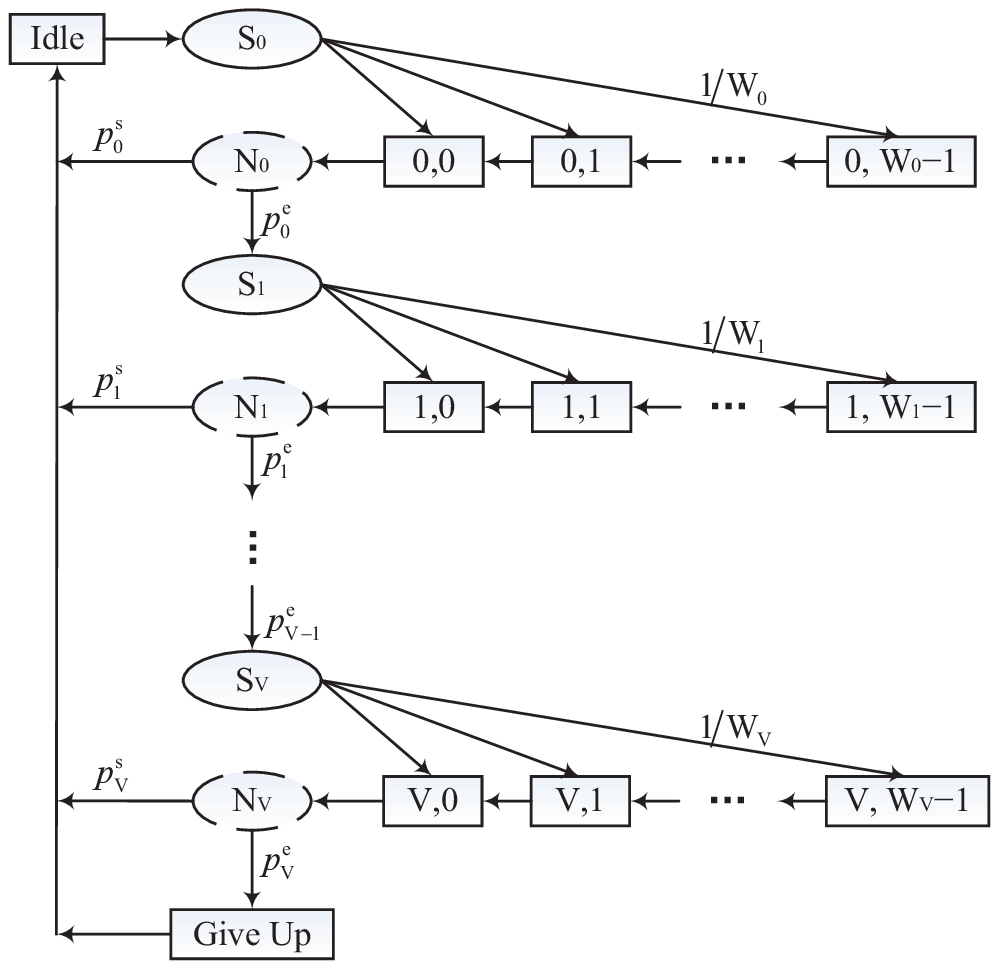}
    \caption{Markov Chain model for the GONORA scheme.}
    \label{Fig_Markov}
\end{figure}

\begin{figure}
    \centering
    \includegraphics[scale=0.75]{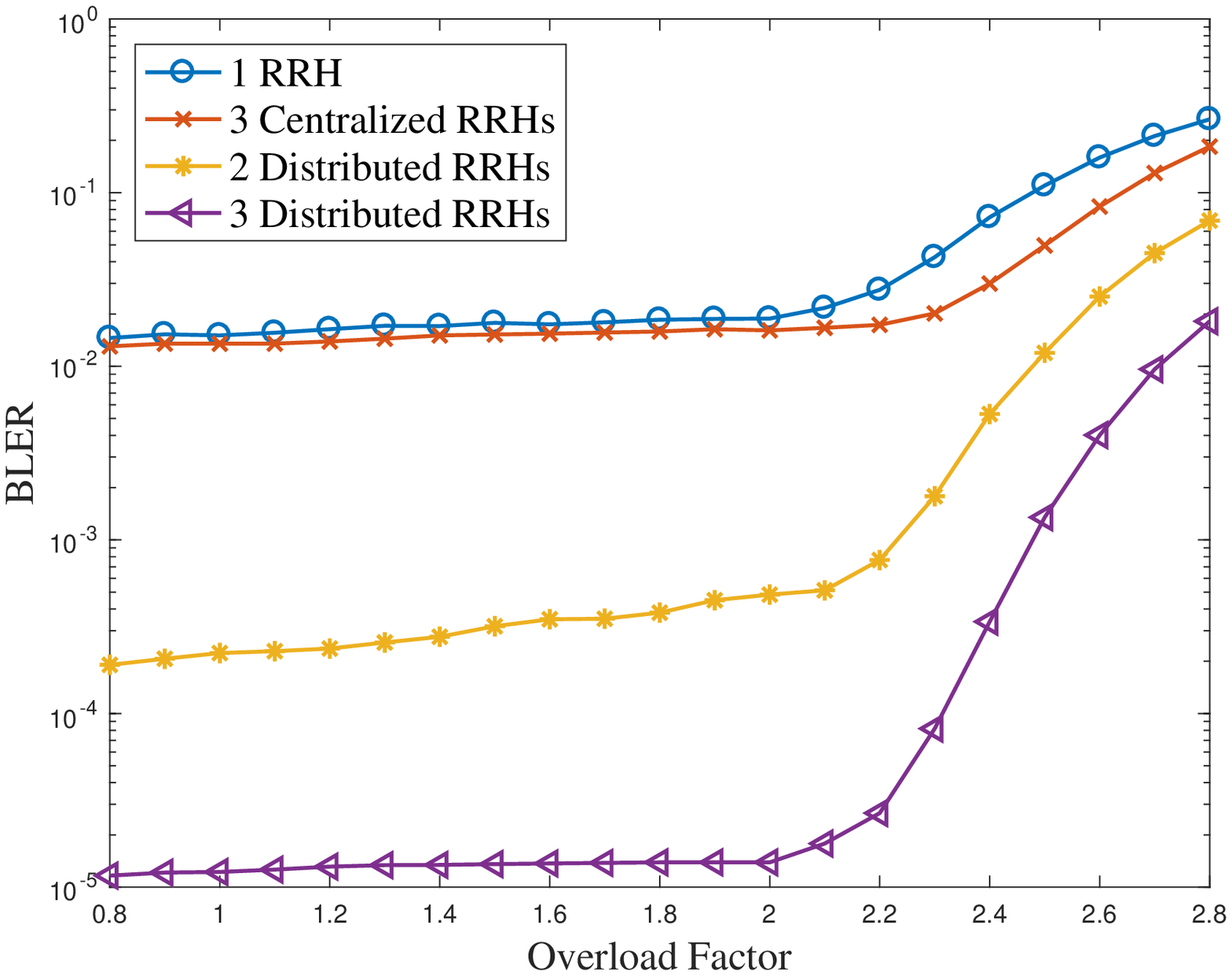}
    \caption{BLER of GONORA with different RRHs.}
    \label{Fig_BLER_RRH}
\end{figure}

\begin{figure}
    \centering
    \includegraphics[scale=0.85]{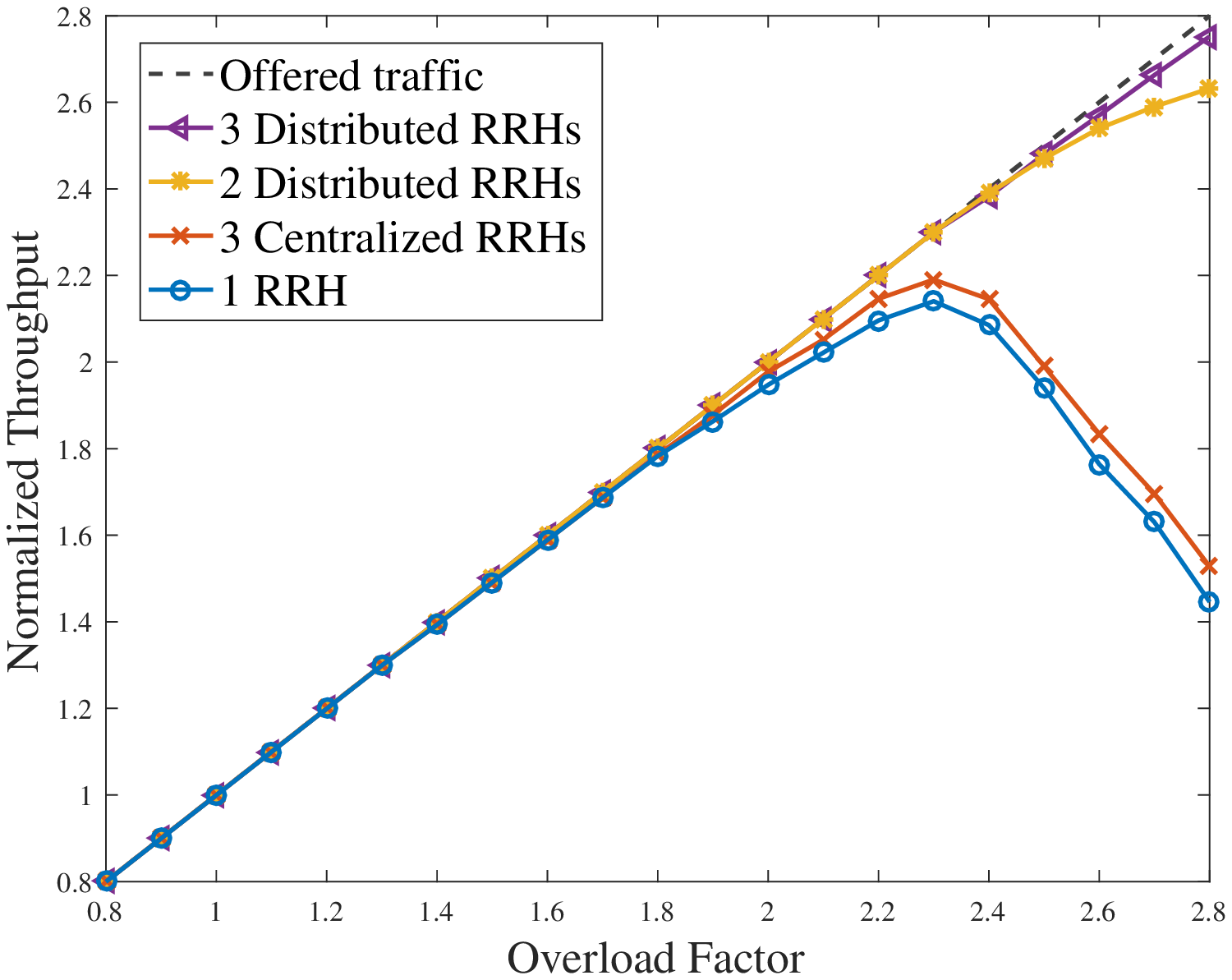}
    \caption{Normalized throughput of GONORA with different RRHs.}
    \label{Fig_Throughput_RRH}
\end{figure}

\begin{figure}
    \centering
    \includegraphics[scale=0.75]{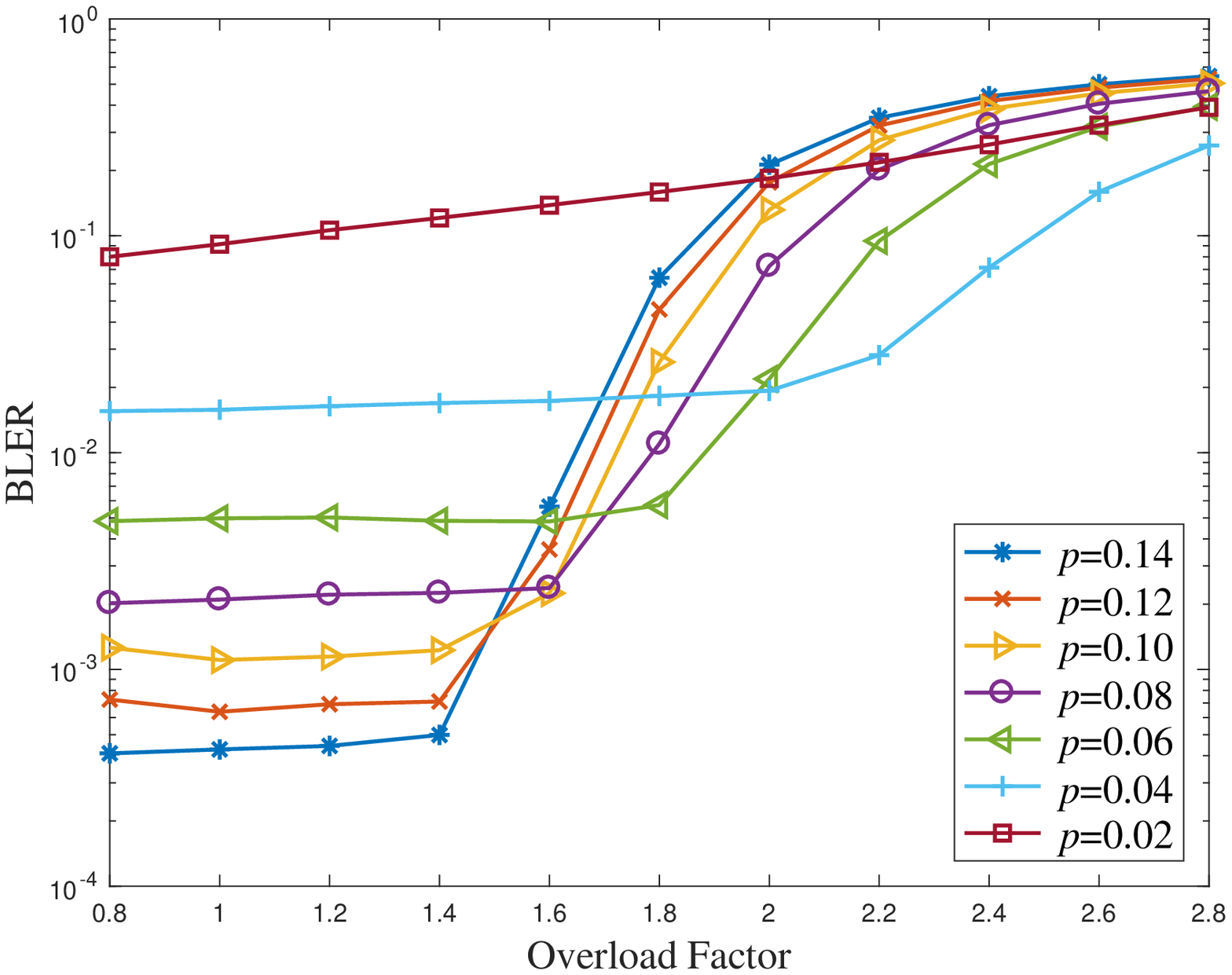}
    \caption{BLER of GONORA with different $p$.}
    \label{Fig_BLER_p}
\end{figure}

In 5G UDHN, the fact that the density of MTC devices will be much higher than that of UEs makes the random access for MTC devices very challenging. To support a massive number of MTC devices, we propose a generalized orthogonal/non-orthogonal random access (GONORA) scheme for 5G UDHN, which is a grant-less transmission scheme.

Let physical resource pool (PRP) denote the basic transmission block in GONORA, where a PRP consists of $\omega$ resource units (RUs). The procedure of GONORA can be modelled as a Markov chain, as shown in Fig. \ref{Fig_Markov}. An MTC device with a new packet to transmit generates a random discrete-time backoff interval before transmitting to minimize the congestion, where the backoff interval is uniformly chosen in the range $(0, \text{W}_v)$, $0\leq v\leq \text V$, with subscript $v$ denoting the repetition times of the new packet. Here, we adopt an exponential backoff scheme \cite{IEEEJSAC2000DCF}, where the content window $\text{W}_v=2^v \text{W}_0$ with $\text{W}_0$ being the initial content window size. When the backoff time counter reaches zero, the MTC device chooses $\text{N}_v$ RUs from the next coming PRP to transmit part of the new packet based on the size of $\text{N}_v$. The value of $\text{N}_v$ is related with the size of the packet, the repetition times, and the payload of PRP. If it is successfully transmitted, the MTC device returns to idle state to wait for transmitting the rest of the new packet or to wait for the next packet. If the MTC device does not receive an acknowledge (ACK) in time, it will prepare for the next repetition. After reaching the maximum allowed repetition times, i.e., $\text V$ in Fig. \ref{Fig_Markov}, the MTC device will give up repetition and return to the idle state no matter whether the packet is successfully transmitted or not.

In this Markov chain, the one-step transition probabilities are
\begin{align}
\left\{ \begin{array}{l}
\Pr \left( {\left. {v,k} \right|v,k + 1} \right) = 1,\;\;\;\;\;k \in \left( {0,\;{{\text{W}}_v} - 2} \right),\;v \in \left( {0,\;{\text{V}}} \right)\\
\Pr \left( {\left. {v,k} \right|{{\text{S}}_v}} \right) = {1 \mathord{\left/
 {\vphantom {1 {{{\text{W}}_v}}}} \right.
 \kern-\nulldelimiterspace} {{{\text{W}}_v}}},\;\;\;\;\;\;k \in \left( {0,\;{{\text{W}}_v} - 1} \right),\;v \in \left( {0,\;{\text{V}}} \right)\\
\Pr \left( {\left. {{{\text{S}}_{v + 1}}} \right|{{\text{N}}_v}} \right) = p_v^{\text{e}},\;\;\;\;\;\;\;\;\;v \in \left( {0,\;{\text{V}} - 1} \right)\\
\Pr \left( {\left. {idle} \right|{{\text{N}}_v}} \right) = p_v^{\text{s}},\;\;\;\;\;\;\;\;\;\;v \in \left( {0,\;{\text{V}}} \right)
\end{array} \right.
\end{align}
where the success and error probabilities $p_v^{\text{s}}$ and $p_v^{\text{e}}$ of the $v$-th repetition are dominated by the channel state and congestion among MTC devices. To simplify the choosing of $\text N_v$, we can choose each RU from one PRP with a probability $p$ with $\text N_v$ following Bernoulli distribution and $E\left[\text N_v\right] = p \omega$. Obviously, $p$ is related with the traffic volume and the size of the PRP, and the value of $p$ can be derived as 
\begin{align}
p = \frac{{\beta \gamma }}{{\sum\limits_{m = 1}^M {{\alpha _m}{\lambda _m}\tau } }}
\end{align}
where $M$ is the number of MTC devices that associated with the same cell, $\alpha_m$ is the average package size of MTC device $m$, $\lambda_m$ is the Poisson parameter, $\tau$ is the time duration of each PRP, $\beta$ is the average capacity of one PRP  without signal superposing, and $\gamma$ denotes the average resource reuse factor. Here, the package arrival of each MTC device is assumed to follow Poisson distribution and, consequently, the average aggregate traffic load during one PRP is ${\sum\nolimits_{m = 1}^M {{\alpha _m}{\lambda _m}\tau } }$.

The block error ratio (BLER) and normalized throughput of GONORA versus overload factor with different RRH numbers are shown in the Figs. \ref{Fig_BLER_RRH} and \ref{Fig_Throughput_RRH}, respectively. Here, the overload factor is defined as the ratio of the MTC device number and the RU number per PRP, and the normalized throughput is defined as the average number of UEs that successfully transmit their packets on one RU. In Fig. \ref{Fig_BLER_RRH}, it is shown that the BLER increases with the increase of overload factor due to the congestion of MTC devices. In Fig. \ref{Fig_Throughput_RRH}, we observe that the normalized throughput approaches the offered traffic with low overload factor, which is due to the fact that the packets are quasi orthogonal in such a scenario. With the increase of overload factor, the orthogonality among the transmitted packets is destroyed, which results in that more RRHs are required to mitigate the congestion. In Fig. \ref{Fig_BLER_p}, the BLER of GONORA versus overload factor with different RU selecting probability $p$ is plotted. We observe that the BLER decreases with $p$ when overload factor is small. Obviously, the higher the value of $p$, the larger the repetition times per packet, and hence the higher the received SINR. With the increase of overload factor, the congestion becomes serious, which leads to higher BLER. 

\section{Challenges of UDHN}
The challenges of the UDHN are from deploying and operating the cellular networks to satisfy the unprecedented mobile device increase and the explosive traffic load growth with limited radio resources. Hence, the most challenge of the UDHN is the interference management. Furthermore, due to the mobility of massive mobile devices, mobility management, mobile association, and channel estimation are also key challenges of the UDHN.

\subsection{Interference Management}
Due to the decrescent distance between neighbouring cells caused by continued network densification, the co-channel interference will become more and more serious and may severely deteriorate the network performance, especially the QoS of cell-edge users. Hence, effective interference management is crucial for guaranteeing the user experience \cite{IEEEWC2016Interf}.

To address the co-channel interference, inter-cell interference coordination (ICIC) has been introduced in LTE networks, where the neighbouring BSs allocate different radio resources to their users in some way to mitigate inter-cell interference (ICI) based on the received interference status of their neighbours from X2 interface. A straightforward way of allocating different resources to cell-edge users is adopting fractional frequency reuse (FFR) strategy, where different frequency reuse factors are applied in the cell center and cell edge regions to mitigate ICI. It is noticed that we can also adopt the concept of FFR to mitigate the co-channel interference among geographically overlapped cells by assigning different frequency resources to different cells in overlapped area. Another effective ICIC strategy in LTE is CoMP, which turns the ICIs into useful signals, especially for cell-edge user, through joint processing or coordinated scheduling/coordinated beamforming.

In 5G UDHN, besides FFR and CoMP, advanced interference management strategies are needed to tackle the more serious interference status. Recently, two new interference management strategies have emerged: interference shaping (IS) and interference exploitation (IE) \cite{IEEEWC2016Interf}. The concept of IS is to linearly combine the interference signals in a certain way to eliminate the aggregated interference effect at receivers, and the representative IS technique is interference alignment (IA), which aligns multiple interfering signals such that the received signal can be projected into the null space of the interfering signals to decode the desired signal with no interference. IE exploits the interference as side information to improve the throughput, such as network coding and index coding.

The abovementioned four interference management strategies are network-side approaches. To further enhance the interference management, user-side approach is another option to alleviate the interference issues in 5G UDHN. In user-side approach, advanced receiver is required to take use of the interference signals structure, and the desired signals can be derived by subtracting the reconstructed interfering signals from the received signals. Joint network-side and user-side interference managements could significantly mitigate the interference.

\subsection{Mobility Management and Mobile Association}
The continued network densification and increased heterogeneity in the UDHN also poses challenges for the mobility management \cite{IEEEJSAC20145G}. A natural way for mobility management in the UDHN is to decouple the user and control planes, where the user plane protocol stack is conducted through small cell to improve the  throughput and the control plane protocol stack is preferred to be conducted by macrocell to reduce the handover frequency. After adopting C-RAN architecture, the mobility management will become more easier with the decoupling of user and control planes, as all necessary information is in the same BBU pool and is easy to be passed from source RRH to target RRH to prepare the handover procedure. Furthermore, the advanced cloud processing could also reduce the intra-BBU pool handover delay. Another way to ease the mobility management in the UDHN is to restrict highly mobile users to macrocells at lower frequencies, thereby forcing them to tolerate lower data rates while minimizing the frequency of the handover.

Most of the existing mobile association schemes are developed based on the downlink channel state information (CSI). In the UDHN, different types of small cells with different transmit powers are densely deployed, where multiple access points are available in both uplink and downlink. Hence, mobile association, which adaptively selects the uplink and downlink access points to guarantee the QoS while balancing the traffic load in different cells, becomes a challenging and fundamental issue. One natural way for mobile association is to select the serving cell with the highest instantaneous receiving power. To reduce the probability of handover, we can also select the serving cell based on long-term CSI. With C-RAN, the mobile association can be performed in the BBU pool to ease the collaboration among cells to guarantee the QoS  from the perspective of the UEs and balance the traffic load from the perspective of the networks.

\subsection{Channel Estimation}
Channel state information is essential not only for resource scheduling and demodulation, but also for coordination among cells in 5G UDHN. And the use of wide channel bandwidths and massive antennas poses significant challenges for the channel estimation. For example, CoMP relies on the availability of CSIs strictly, and fast CSI feedback is required to tackle the channel changes. To alleviate the CSI estimation and signalling overhead, instead of allowing coordination among all the cells, the cells can be clustered so that only certain cells in the same cluster joint the CoMP. In IA algorithm, global and instantaneous CSI at transmitter is also crucial for aligning and removing the interference. 

One simple method of reducing the CSI overhead is to use the channel stochastics to replace the instantaneous CSI, e.g., using the channel autocorrelation function instead of CSI itself in blind IA, at the cost of performance degradation. The second method is to adopt the channel reciprocal properties to estimate the uplink (downlink) channel based on the downlink (uplink) signals to avoid the reference signals and to reduce the feedback overhead. The second method is attractive for MIMO channel, especially for massive MIMO channel. However, the channel reciprocality only exists in time division duplexing systems, and the channel reciprocal is also unavailable for DUDe.

\section{Conclusions}
Network architecture dominates the performance of the networks. This article provides a potential network architecture for 5G to satisfy the 1000x traffic increase. The future cellular networks will be an ultra-dense multi-tier heterogeneous networks along with D2D pairs and MTC devices. To accommodate different tier cells, DUDe and C-RAN will be introduced for 5G. We have proposed the GONORA scheme for massive MTC devices and analyzed the challenges of the UDHN. Nevertheless, there are still quite a number of outstanding problems that need further investigations for 5G.

\bibliographystyle{IEEEtran}

\begin{IEEEbiographynophoto}
{Jianping An}  received the Ph.D. degree from Beijing Institute of Technology, Beijing, China, in 1996. He joined the School of Information and Electronics, Beijing Institute of Technology in 1995, where he is now a full Professor. He is currently the Dean of the School of Information and Electronics, Beijing Institute of Technology, Beijing. His research interests are in the field of digital signal processing, cognitive radio, wireless networks, and high-dynamic broadband wireless transmission technology.
\end{IEEEbiographynophoto}

\begin{IEEEbiographynophoto}
{Kai Yang} [M'12] received the B.E. and Ph.D. degrees from National University of Defense Technology and Beijing Institute of Technology, China, in 2005 and 2010, respectively, both in communications engineering. Now, he is with the School of Information and Electronics, Beijing Institute of Technology, Beijing, China. His current research interests include convex optimization, cooperative communications, MIMO systems, resource allocation, and interference mitigation.
\end{IEEEbiographynophoto}

\begin{IEEEbiographynophoto}
{Jinsong Wu} [SM] is elected Vice-Chair of Technical Activities, IEEE Environmental Engineering Initiative (EEI), a pan-IEEE effort under IEEE Technical Activities Board (TAB). He is founder and founding Chair of IEEE Technical Committee on Green Communications and Computing (TCGCC). One of his papers has been selected as 2017 IEEE Systems Journal Best Paper Award. He was the leading Editor and co-author of the comprehensive book Green Communications: Theoretical Fundamentals, Algorithms, and Applications (CRC Press, 2012).
\end{IEEEbiographynophoto}

\begin{IEEEbiographynophoto}
{Neng Ye}  received the B.S. degree with honor in 2015, and is currently working toward his Ph.D. degree in electronic engineering at Beijing Institute of Technology, China. In his work, he focuses on both research and standardization of next generation radio access. His research interests include information theory, non-orthogonal multiple access, channel coding, resource allocation, physical layer security, and machine learning.
\end{IEEEbiographynophoto}

\begin{IEEEbiographynophoto}
{Song Guo} [SM] is a full professor at the Department of Computing, The Hong Kong Polytechnic University. He has published over 300 papers in refereed journals/conferences and received multiple IEEE/ACM best paper awards. Dr. Guo is an editor of the IEEE Transactions on Green Communications and Networking and the Secretary of the IEEE Technical Committee on Big Data. He is a Senior Member of the ACM and an IEEE Communications Society Distinguished Lecturer.
\end{IEEEbiographynophoto}

\begin{IEEEbiographynophoto}
{Zhifang Liao} received the B.S. degree in Changsha Railway University, China, in 1990 and PhD in School of Information Science and Engineering, Central South University, China, in 2008. She currently works as an associate professor in the School of Software, Central South University, Changsha, China. Her research interests include data mining, open source software, and social networks. She has won Excellent Teaching Quality Award and 2nd Prize of Bilingual Teaching Competition of Central South University.
\end{IEEEbiographynophoto}

\setcounter{figure}{0}
\newpage
\begin{figure}
    \centering
    \includegraphics[scale=1]{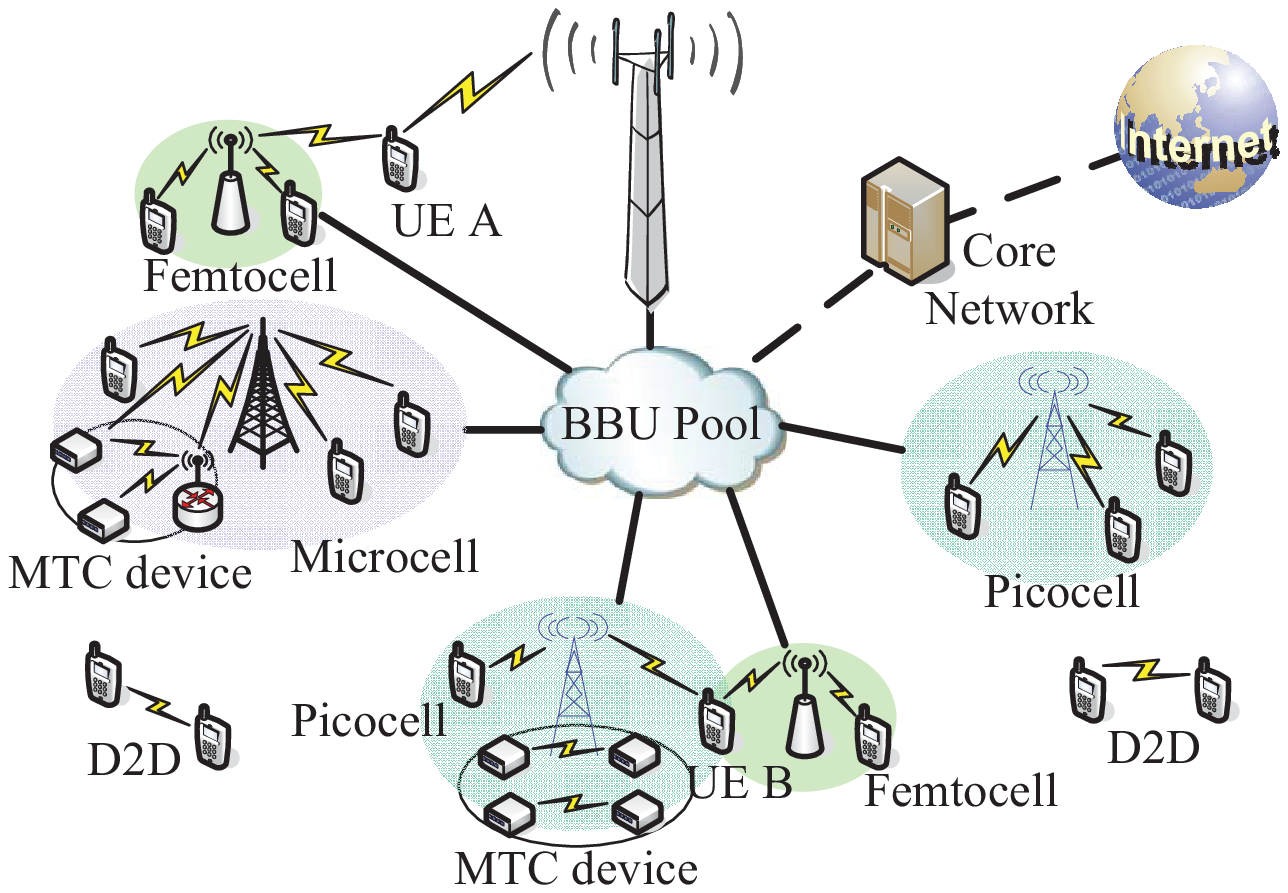}
    \caption{The proposed potential network architecture for 5G ultra-dense heterogeneous networks.}
    \label{Fig_Architecture}
\end{figure}

\newpage

\begin{figure}
    \centering
    \includegraphics[scale=1.2]{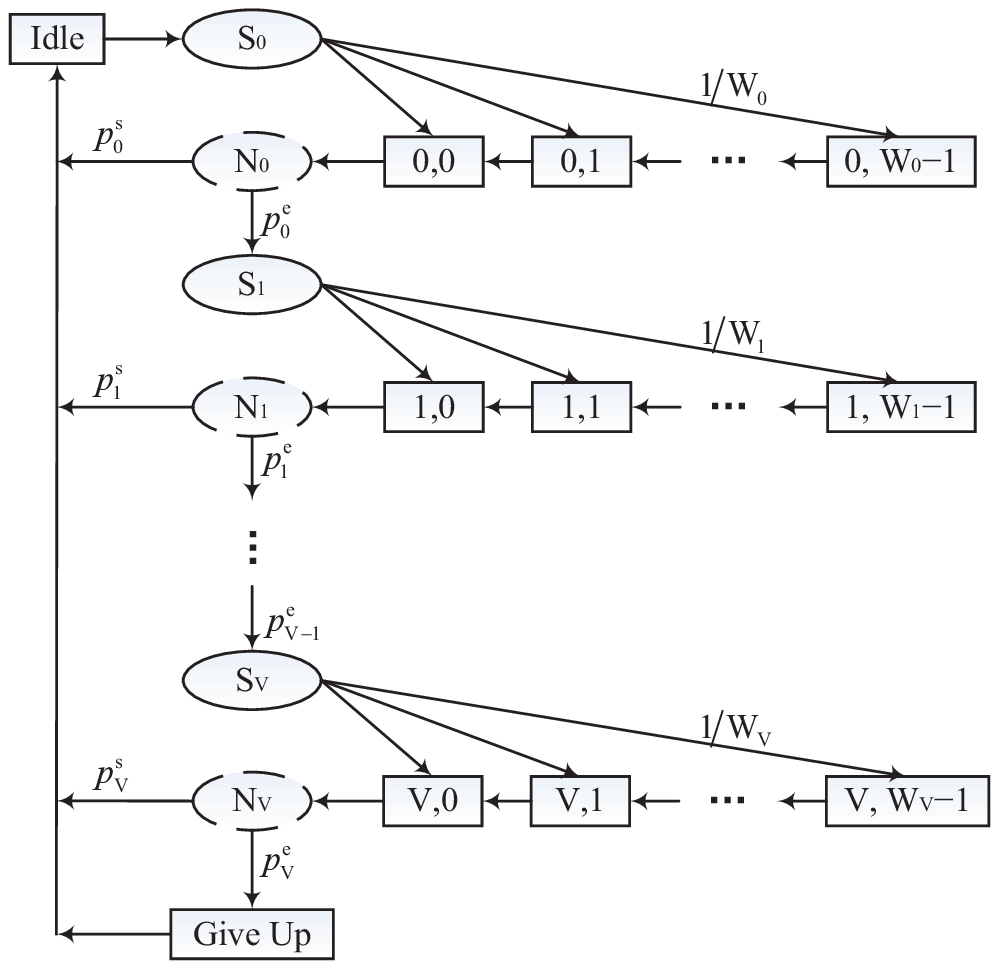}
    \caption{Markov Chain model for the GONORA scheme.}
    \label{Fig_Markov}
\end{figure}

\newpage
\begin{figure}
    \centering
    \includegraphics[scale=0.75]{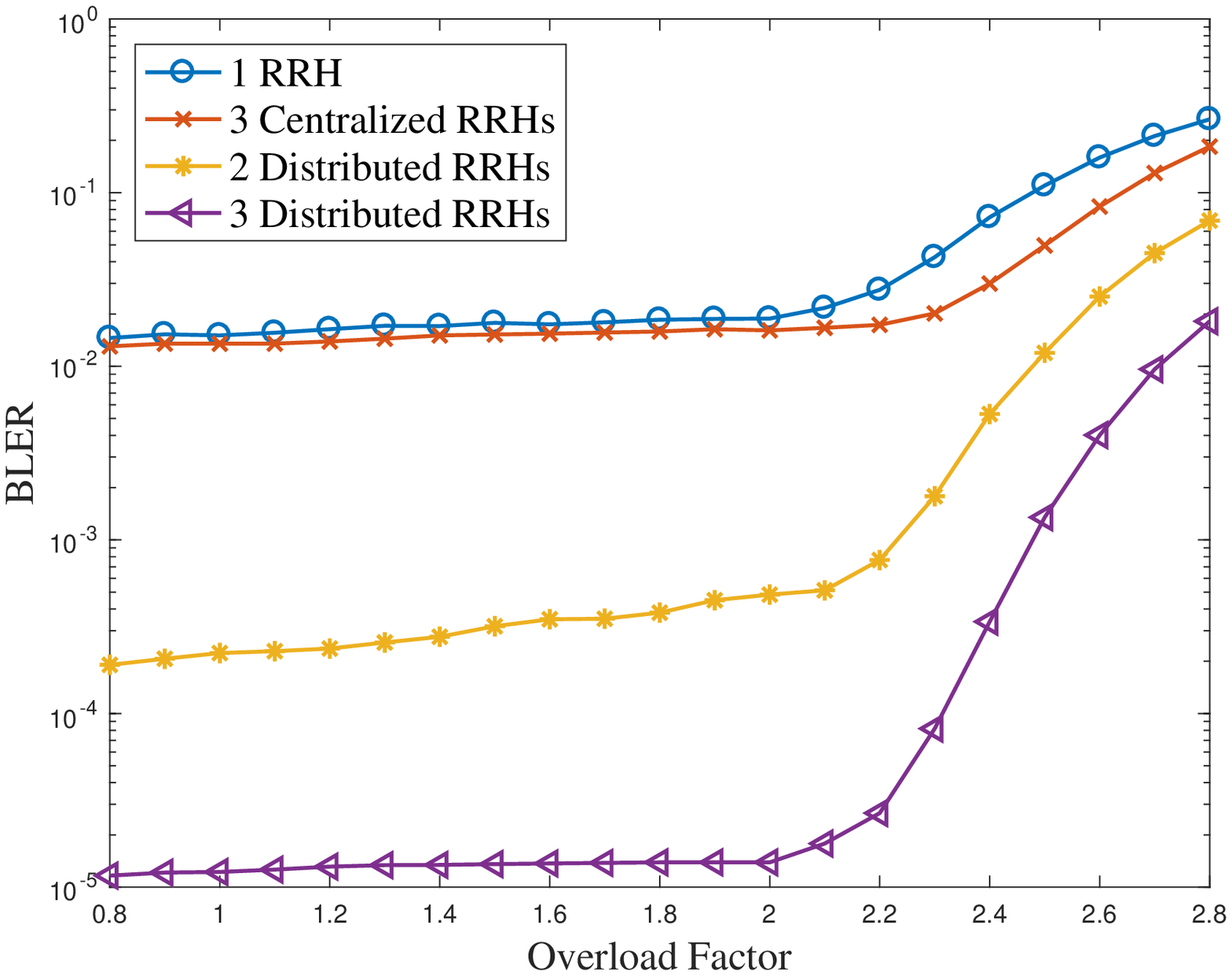}
    \caption{BLER of GONORA with different RRHs.}
    \label{Fig_BLER_RRH}
\end{figure}

\newpage
\begin{figure}
    \centering
    \includegraphics[scale=0.85]{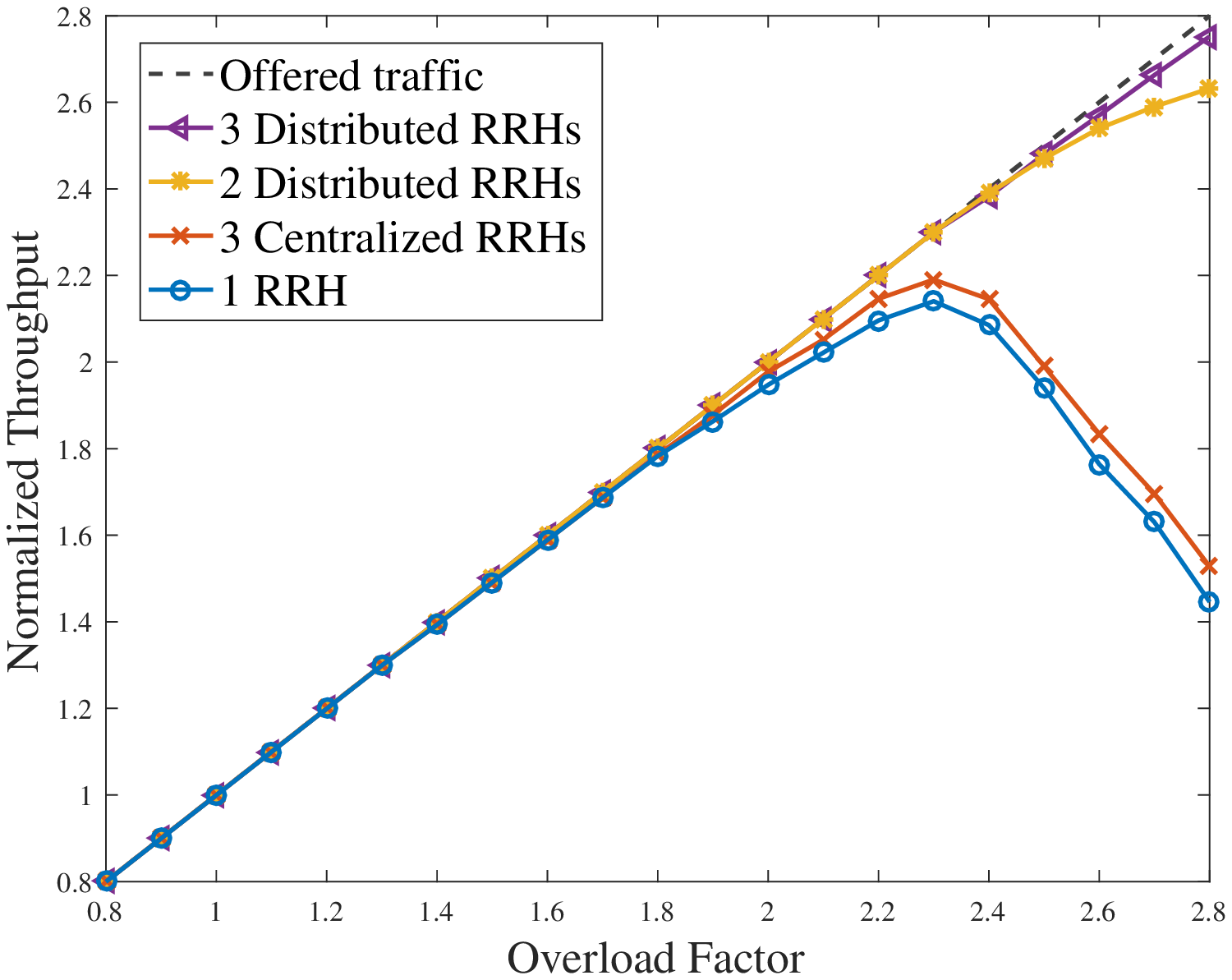}
    \caption{Normalized throughput of GONORA with different RRHs.}
    \label{Fig_Throughput_RRH}
\end{figure}

\newpage
\begin{figure}
    \centering
    \includegraphics[scale=0.75]{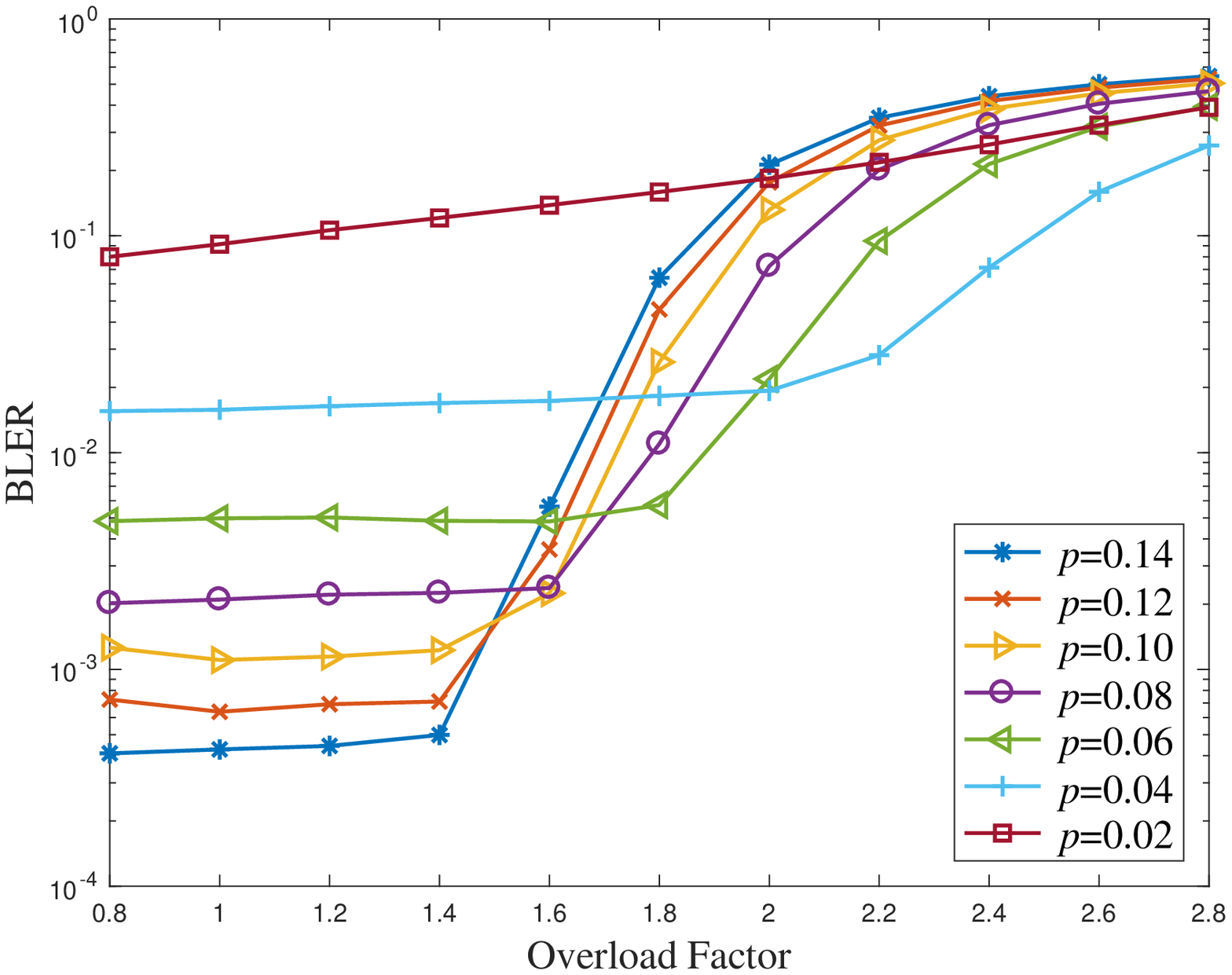}
    \caption{BLER of GONORA with different $p$.}
    \label{Fig_BLER_p}
\end{figure}

\end{document}